\documentclass[conference,a4paper]{IEEEtran}

\usepackage{amsmath}
\usepackage{amssymb}
\usepackage{latexsym}
\usepackage{array,arydshln}
\usepackage{multirow}
\usepackage{graphicx}
\usepackage{float}
\usepackage{bm}
\usepackage{breqn}
\usepackage{ctable}
\usepackage{cite} 

\newtheorem{definition}{Definition}
\newtheorem{theorem}{Theorem}
\newtheorem{lemma}{Lemma}
\newtheorem{proposition}{Proposition}

\newcommand{\bs}[1]{\ensuremath{\boldsymbol{#1}}}

\newcommand{\beq}{\begin{equation}}
\newcommand{\enq}{\end{equation}}
\newcommand{\beal}{\begin{align*}}
\newcommand{\enal}{\end{align*}}

\newcommand{\fs}[1]{\ensuremath{f_{\mathrm{#1},\mathrm{s}}}}
\newcommand{\fp}[1]{\ensuremath{f_{\mathrm{#1},\mathrm{p}}}}

\newcommand{\xsa}[2]{\ensuremath{p_{\mathrm{#1},\mathrm{#2}}}}
\newcommand{\ysa}[2]{\ensuremath{p_{\mathrm{#1},\mathrm{#2}}}}

\newtheorem{example}{Example}

\IEEEoverridecommandlockouts

\begin{document}

\title{Threshold Saturation for Spatially Coupled Turbo-like Codes over the
  Binary Erasure Channel}


\author{
\IEEEauthorblockN{Saeedeh Moloudi$^\dag$, Michael Lentmaier$^\dag$, and Alexandre Graell i Amat$^\ddag$}
\IEEEauthorblockA{$\dag$Department of Electrical and Information
  Technology, Lund University, Lund, Sweden \\
  $\ddag$Department of Signals and Systems, Chalmers University of Technology, Gothenburg, Sweden}\\
              \thanks{This work was supported in part by the Swedish Research Council (VR) under grant \#621-2013-5477.}\vspace*{-1cm}
}


\maketitle

\begin{abstract}
In this paper we prove threshold saturation for spatially coupled turbo codes (SC-TCs) and braided convolutional codes (BCCs) over the binary erasure channel.
We introduce a compact graph representation for the ensembles of SC-TC and BCC codes which simplifies their description and the analysis of the message passing decoding.
We demonstrate that by few assumptions in the ensembles of these codes, it is possible to rewrite their vector recursions in a form which places these ensembles under the category of scalar admissible systems. 
This allows us to define \emph{potential functions} and prove threshold saturation using the proof technique introduced by Yedla {\em et al.}. 
\end{abstract}

\IEEEpeerreviewmaketitle

\section{Introduction}


Low-density parity-check (LDPC) convolutional codes \cite{JimenezLDPCCC}, also known as spatially coupled LDPC (SC-LDPC) codes \cite{Kudekar_ThresholdSaturation} have received a great deal of attention in the recent years as a result of their excellent performance under iterative decoding. 
In particular, it has been shown that the threshold of a belief propagation (BP) decoder improves to the threshold of an optimal maximum a-posteriori (MAP) decoder. 
This remarkable phenomenon is called threshold saturation.

Spatial coupling is not limited to LDPC codes. 
Braided convolutional codes (BCCs) are a class of spatially coupled (SC) codes introduced in \cite{ZhangBCC}.
Recently, the authors introduced spatially coupled turbo codes (SC-TCs) \cite{MoloudiISTC14}, as the SC counterparts of parallel \cite{BerrouTC}  and serially \cite{Benedetto98Serial} concatenated convolutional codes.
In \cite{MoloudiISTC14,MoloudiISIT14,Moloudi_SPCOM14,Moloudi_ISTW14}, we investigated threshold saturation for SC parallel concatenated codes (SC-PCCs), SC serially concatenated codes (SC-SCCs) and BCCs over the binary erasure channel (BEC).
We derived closed-form density evolution (DE) equations for SC-TCs and BCCs and investigated their decoding thresholds.
Our numerical results suggest that threshold saturation occurs for SC-PCCs, SC-SCCs and BCCs. 

In this paper, we formally prove threshold saturation for SC-TCs and BCCs over the BEC.
Our proof relies on the proof technique based on potential functions, recently proposed by Yedla \textit{et al.} \cite{Yedla2012,Yedla2012Vector}.
We introduce a compact graph representation to describe PCC, SCC and BCC ensembles.
Similar to a protograph \cite{Loeliger}, the compact graph makes it easier to illustrate the analysis of the message passing decoding. 
We then demonstrate that by few assumptions, the DE recursions of SC-TCs \cite{MoloudiISTC14,Moloudi_SPCOM14} and BCCs \cite{MoloudiISIT14,Moloudi_SPCOM14}, can be rewritten in a form that corresponds to the recursion of a scalar admissible system as in \cite{Yedla2012}.
This makes it possible to derive suitable potential functions for TCs and uncoupled BCCs.
Finally, we prove threshold saturation for SC-TCs and BCCs following the same lines as the proof in \cite{Yedla2012} for SC-LDPC codes.
\section{Compact Graph Representation} \label{CGR}
It is possible to analyze message passing decoding algorithms in an efficient way by the use of factor graphs \cite{Loeliger}.
However the conventional factor graph of codes with convolutional component codes, such as PCCs, SCCs and BCCs, gets very large as the length of the component codes increases.
In this section we introduce a more compact graph representation, in which each trellis is represented by a single factor node and each collection of variables of the same type is represented by a single variable node.
We use this compact graph representation to obtain DE equations and describe the spatially coupled ensembles.

\begin{figure}[t]
  \centering
    \includegraphics[width=0.78\linewidth]{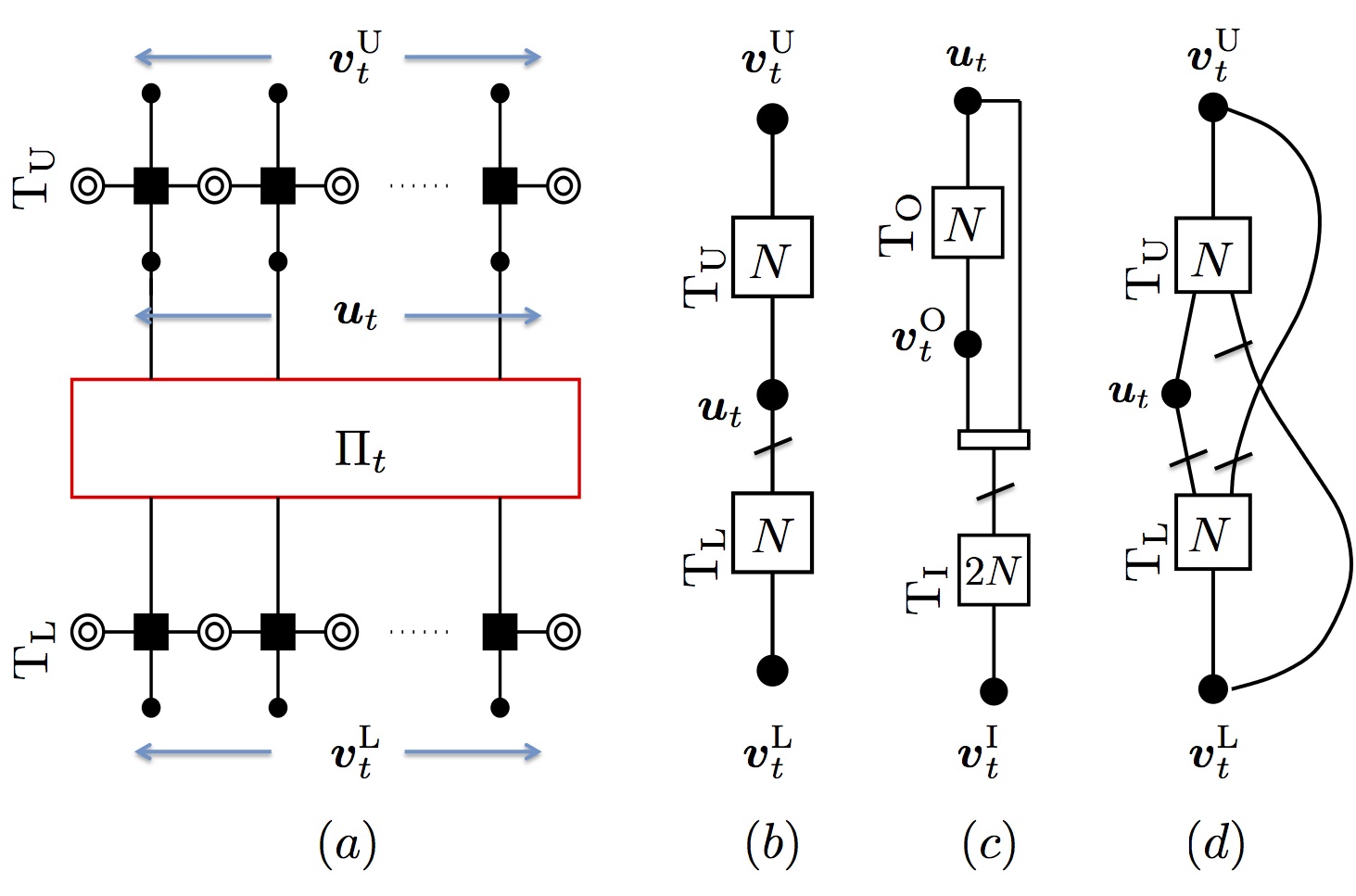}
\vspace{-2ex}
\caption{(a) Conventional factor graph of a PCC, compact graph of a (b) PCC, (c) SCC, (d) BCC.}
\label{Fig1}
\vspace{-4ex}
\end{figure}

Fig.~\ref{Fig1}$(a)$ shows the conventional factor graph representation of a rate $R = 1/3$ PCC.
This code is built from two rate-$1/2$ recursive systematic encoders, referred to as upper and lower encoders; we call the corresponding trellises upper and lower trellises and denote them by $\text{T}_{\text{U}}$ and $\text{T}_{\text{L}}$, respectively.
The information sequence at time slot $t$ is denoted by $\boldsymbol{u}_t$ and is a vector of $N$ bits $\boldsymbol{u}_t=(u_1,u_2,\dots,u_{N})$.
The information sequence $\boldsymbol{u}_t$ and its reordered copy are encoded by the upper and lower encoder to produce parity sequences  $\boldsymbol{v}_t^{\text{U}}$ and $\boldsymbol{v}_t^{\text{L}}$, respectively. 

Fig.~\ref{Fig1}$(b)$ shows the compact graph representation of this code.   
Each of the sequences $\boldsymbol{u}_t$, $\boldsymbol{v}_t^{\text{U}}$ and $\boldsymbol{v}_t^{\text{L}}$ are represented by a single black circle (variable node).
Thus, each variable node in the compact graph corresponds to a number of code symbols.
The trellises are replaced by squares (factor nodes) which are labeled by their length\footnote{The length of a trellis is equal to the length of each of the sequences which are connected to that trellis.}.
The permutation is represented by a line that crosses the edge which connects $\boldsymbol{u}_t$ to $\text{T}_{\text{L}}$ in order to emphasize that a reordered copy of $\boldsymbol{u}_t$ is used in $\text{T}_{\text{L}}$.
The transmitted code sequence is $\boldsymbol{v}=(\boldsymbol{u}_t,\boldsymbol{v}_t^{\text{U}},\boldsymbol{v}_t^{\text{L}})$.

Fig.~\ref{Fig1}$(c)$ depicts the compact graph representation of a rate $R=1/4$ SCC built from two rate-$1/2$ recursive systematic convolutional encoders, referred to as inner and outer encoders.
We call the corresponding trellises inner and outer trellises and denote them by $\text{T}_{\text{I}}$ and $\text{T}_{\text{O}}$, respectively.
At time $t$, the information sequence $\boldsymbol{u}_t$, of length $N$, is encoded by the outer encoder to produce the parity sequence $\boldsymbol{v}_t^{\text{O}}$.
Then, $\boldsymbol{u}_t$ and $\boldsymbol{v}_t^{\text{O}}$ are multiplexed and reordered to create a sequence $\tilde{\boldsymbol{v}}_t^{\text{O}}$ whose length is $2N$. 
The sequence $\tilde{\boldsymbol{v}}_t^{\text{O}}$ is encoded by the inner encoder to produce the parity sequence $\boldsymbol{v}_t^{\text{I}}$. 
The transmitted code sequence is $\boldsymbol{v}=(\boldsymbol{u}_t,\boldsymbol{v}_t^{\text{O}},\boldsymbol{v}_t^{\text{I}})$.

BCCs consist of two rate-2/3 component convolutional encoders.
As for PCCs, we call the component encoders upper and lower and the corresponding trellises as upper and lower trellises and denote them by $\text{T}_{\text{U}}$ and $\text{T}_{\text{L}}$, respectively.
BCCs are inherently SC but we can define uncoupled BCCs by tailbiting a chain of a coupled code with coupling length $L=1$.
Fig.~ \ref{Fig1}$(d)$ shows the compact graph of uncoupled BCCs. 
At time $t$, the parity sequences of $\text{T}_{\text{U}}$ and $\text{T}_{\text{L}}$ are denoted by $\boldsymbol{v}_t^{\text{U}}$ and $\boldsymbol{v}_t^{\text{L}}$, respectively.
The information sequence $\boldsymbol{u}_t$ and a reordered copy of $\boldsymbol{v}_t^{\text{L}}$ are used in $\text{T}_{\text{U}}$ to produce $\boldsymbol{v}_t^{\text{U}}$.
Likewise, some reordered copies of  $\boldsymbol{u}_t$ and $\boldsymbol{v}_t^{\text{U}}$ are used in $\text{T}_{\text{L}}$ to produce $\boldsymbol{v}_t^{\text{L}}$.
The transmitted code sequence is $\boldsymbol{v}=(\boldsymbol{u}_t,\boldsymbol{v}_t^{\text{U}},\boldsymbol{v}_t^{\text{L}})$.
\section{Density Evolution and Scalar Admissible System} \label{Sec1}
In this section, we first define a scalar admissible system.
Then we show that by few assumptions in the ensembles of TCs and BCCs, it is possible to rewrite their DE recursions in a form which corresponds to the recursion of a scalar admissible system.
\begin{definition}[\cite{Yedla2012}]\label{def1} A scalar admissible system $(f,g)$, is defined by the recursion
\begin{equation}
\label{recursion}
x^{(i)}=f\Big( g(x^{(i-1)});\varepsilon\Big),
\end{equation}
where $f : [0,1] \times [0,1] \rightarrow [0,1]$ and $g : [0,1] \rightarrow [0,1]$ satisfy the following conditions.
\begin{itemize}
\item $f$ is increasing in both arguments $x,\varepsilon \in (0,1]$; 
\item $g$ is increasing in $x \in (0,1]$; 
\item $f(0;\varepsilon)=f(x;0)=g(0)=0$;
\item $f$ and $g$ have continuous second derivatives.
\end{itemize}
\end{definition}

\subsection{Parallel Concatenated Codes} \label{PCCs}
Consider the PCC in Fig.~\ref{Fig1}$(b)$. 
To formulate the DE equations as we obtained in \cite{MoloudiISTC14}, let $p_{\text{U},\text{s}}^{(i)}$ and $p_{\text{L},\text{s}}^{(i)}$ denote the extrinsic erasure probabilities to $\boldsymbol{u}_t$ from $\text{T}_{\text{U}}$ and $\text{T}_{\text{L}}$, respectively.
Consider transmission over the BEC with erasure probability $\varepsilon$.
The erasure probabilities to $\text{T}_{\text{U}}$ from $\boldsymbol{u}_t$ and $\boldsymbol{v}_t^{\text{U}}$, in the $i$th iteration, are $\varepsilon \cdot p_{\text{L},\text{s}}^{(i-1)}$ and $\varepsilon$, respectively. 
Thus the DE update for $\text{T}_{\text{U}}$ is given by  
\begin{align}
\label{DEPCC1}
p_{\text{U},\text{s}}^{(i)}=\fs{U}\left(
q_{\text{L}}^{(i-1)},\varepsilon\right),
\end{align}
where
\begin{equation}
\label{DEPCC2}
q_{\text{L}}^{(i-1)}=\varepsilon \cdot p_{\text{L},\text{s}}^{(i-1)}
\end{equation}
 and $\fs{U}$ denotes the transfer function of $\text{T}_{\text{U}}$ for the information bits. 
Similarly, the DE update for $\text{T}_{\text{L}}$ can be written as
\begin{align}
\label{DEPCC3}
p_{\text{L},\text{s}}^{(i)}=\fs{L}\left(
q_{\text{U}}^{(i-1)},\varepsilon\right)
\end{align}
where
\begin{equation}
\label{DEPCC4}
q_{\text{U}}^{(i-1)}=\varepsilon \cdot p_{\text{U},\text{s}}^{(i-1)}.
\end{equation}
The DE equations for PCCs in \eqref{DEPCC1}-\eqref{DEPCC4} involve different edges and hence form a vector recursion.
However, considering identical $\text{T}_{\text{U}}$ and $\text{T}_{\text{L}}$ nodes (i.e., identical component encoders), it follows that $f_{\text{U},\text{s}}=f_{\text{L},\text{s}}\triangleq f_{\text{s}}$. 
Therefore, $p_{\text{U},\text{s}}^{(i)}=p_{\text{L},\text{s}}^{(i)}\triangleq x^{(i)}$. 
Now, using this and by substituting (\ref{DEPCC2}) into (\ref{DEPCC1}) and (\ref{DEPCC3}) into (\ref{DEPCC4}), the DE can be written as a scalar recursion,
\begin{equation}
\label{DEPCC3}
x^{(i)}=f_{\text{s}}(\varepsilon x^{(i-1)},\varepsilon  ),
\end{equation}
where the initial condition is $x^{(0)} = 1$.
Consider $f(x;\varepsilon)=f_{\text{s}}(\varepsilon\cdot x, \varepsilon)$ and $g(x)=x$. 
We show in the following that these two functions meet the conditions in Definition \ref{def1}. Therefore, the recursion \eqref{DEPCC3} is a recursion of a scalar admissible system. 
The function $g(x)=x$ is a simple function and it is easy to show that it satisfies all conditions in Definition \ref{def1}. 
$f(x;\varepsilon)$ is a transfer function of a BCJR decoder with a-priori information $x$ and channel erasure probability $\varepsilon$. 
In the following lemma we show that it satisfies the conditions in Definition \ref{def1}.

\begin{lemma}
\label{remark1}
Consider a terminated convolutional code where all distinct input
sequences have distinct coded sequences. 
For such a system, the transfer function $f(x;\varepsilon)$ of a BCJR decoder with a-priori probability $x$ and channel erasure probability $\varepsilon$, or any convex combination of such transfer functions, satisfies all conditions in Definition \ref{def1}.
\end{lemma}
\begin{IEEEproof}
The BCJR decoder is an optimal APP decoder. Consider two BECs with
erasure probabilities $\varepsilon_1$ and $\varepsilon_2$, with
$\varepsilon_1<\varepsilon_2$.
The BEC with erasure rate $\varepsilon_2$, can be seen as the concatenation
of two BECs with erasure rates $\varepsilon_1$ and $\varepsilon'$,
where $\varepsilon'=1-\frac{1-\varepsilon_2}{1-\varepsilon_1}$.
The data processing inequality implies that $f(x;\varepsilon_1)<f(x;\varepsilon_2)$. This means that the erasure probability at the output of the BCJR decoder is monotone and increases with $\varepsilon$. When $\varepsilon=0$, the input sequence can be recovered perfectly
from the received sequence, as there is a one-to-one mapping of input sequences to coded
sequences. This means $f(x;0)=0$. It is also possible to proof that  $f(x_1;\varepsilon)<f(x_2;\varepsilon)$ for $x_1<x_2$ (not shown due to lack of space).

Finally, $f(x;\varepsilon)$ is a rational function and its poles are outside the interval $x,\varepsilon \in [0,1]$ (otherwise we may get infinite output erasure probability for a finite input erasure probability), so it has continuous first and second derivatives in the interval $x,\varepsilon \in [0,1]$. 
\end{IEEEproof}
\subsection{ Serially Concatenated Codes}
Consider the SCC in Fig.~\ref{Fig1}$(c)$.     
We define by $\xsa{I}{s}^{(i)} $ the extrinsic erasure probability from $\text{T}_{\text{I}}$ to $\boldsymbol{u}_t$ and $\boldsymbol{v}_t^{\text{O}}$.
Likewise, let $\xsa{O}{s}^{(i)}$ and $\xsa{O}{p}^{(i)}$ denote the extrinsic erasure probabilities from $\text{T}_{\text{O}}$ to $\boldsymbol{u}_t$ and $\boldsymbol{v}_t^{\text{O}}$ in the $i$th iteration, respectively. 
Consider the transmission over the BEC with erasure probability $\varepsilon$.
The erasure probabilities from $\boldsymbol{u}_t$ and $\boldsymbol{v}_t^{\text{O}}$ to $\text{T}_{\text{O}}$ in the $i$th iteration both are equal to
\begin{align}
\label{DESCC1}
q_{\text{I}}^{(i-1)}=\varepsilon \cdot \xsa{I}{s}^{(i-1)}.
\end{align}
Thus, the DE equations for $\text{T}_{\text{O}}$ can be written as
\begin{align}
\label{eq:OuterUpdates}
\ysa{O}{s}^{(i)}&=\fs{O}\left(q_{\text{I}}^{(i-1)},q_{\text{I}}^{(i-1)}\right)\\
\label{eq:OuterUpdatep}
\ysa{O}{p}^{(i)}&=\fp{O}\left(q_{\text{I}}^{(i-1)},q_{\text{I}}^{(i-1)}\right),
\end{align}
where $\fs{O}$ and $\fp{O}$ denote the transfer functions of $\text{T}_{\text{O}}$ for the input and parity bits, respectively. 

The input sequence  of the inner encoder consists of $\boldsymbol{u}_t$ and $\boldsymbol{v}_t^{\text{O}}$, so that the erasure probability $q_{\text{O}}^{(i)}$ that comes to $\text{T}_{\text{O}}$ through the set of these two variable nodes is the average of the extrinsic erasure probabilities from $\boldsymbol{u}_t$ and $\boldsymbol{v}_t^{\text{O}}$, i.e.,
\begin{equation}
\label{eq:InnerUpdate3}
q_{\text{O}}^{(i)}=\varepsilon \cdot\frac{\ysa{O}{s}^{(i)}+\ysa{O}{p}^{(i)}}{2} .
\end{equation}
Let  $\fs{I}$  denote the transfer function of $\text{T}_{\text{I}}$ for the input bits.
 The DE equations for $\text{T}_{\text{O}}$ can be written as
\begin{align}
\label{eq:LowerUpdates}
\ysa{I}{s}^{(i+1)}=\fs{I}\left(
q_{\text{O}}^{(i)},\varepsilon\right).
\end{align}
Equations (\ref{DESCC1}) to (\ref{eq:LowerUpdates}) show that the DE for SCCs in Fig. \ref{Fig1} (c) is a vector recursion.
However, for identical $\text{T}_{\text{O}}$ and $\text{T}_{\text{I}}$, it follows $\fs{I}=\fs{O}\triangleq f_{\text{s}}$ and $\fp{I}=\fp{O}\triangleq f_{\text{p}}$. 
Using this and and $q_{\text{I}}^{(i-1)}\triangleq x^{(i)}$, by substituting  \eqref{eq:OuterUpdates}-\eqref{eq:LowerUpdates} into \eqref{DESCC1}, the DE recursion can be written as 
\begin{equation}
\label{eq:SCCrec}
x^{(i+1)}=\varepsilon \cdot f_{\text{s}}\Big(\varepsilon g(x^{(i)}),\varepsilon\Big),
\end{equation}
where
\begin{equation}
\label{eq:gSCC}
g(x^{(i)})=\frac{f_{\text{s}}\Big(x^{(i)},x^{(i)}\Big)+f_{\text{p}}\Big(x^{(i)},x^{(i)}\Big)}{2},
\end{equation}
and the initial condition is $x^{(0)}=1$.

Consider $f(x;\varepsilon)=\varepsilon \cdot f_{\text{s}}(x, \varepsilon)$ and
\[
g(x)=\frac{f_{\text{s}}(x,x)+f_{\text{p}}(x,x)}{2}.
\]
According to Lemma \ref{remark1}, these two functions meet the conditions in Definition \ref{def1} and we can conclude that the DE recursion of SCCs in (\ref{eq:SCCrec}) is a recursion of a scalar
admissible system.

\subsection{Braided Convolutional Codes }\label{BCC}
Consider the BCC in Fig.~\ref{Fig1}$(d)$.  
Let $p_{\text{U},k}^{(i)}$ and $p_{\text{U},k}^{(i)}$, $k=1,2,3$, denote the extrinsic erasure probabilities from $\text{T}_{\text{U}}$ and $\text{T}_{\text{L}}$ in the $i$th iteration, through their $k$th connected edge, respectively. 
The exact DE equations can be written as \cite{MoloudiISIT14}
\begin{align}
p_{\text{U},1}^{(i)}=&f_{\text{U},1}\left(\varepsilon \cdot p_{\text{L},1}^{(i-1)} ,\varepsilon \cdot p_{\text{L},3}^{(i-1)},\varepsilon \cdot p_{\text{L},2}^{(i-1)}\right) \label{eqDE1}\\
p_{\text{U},2}^{(i)}=&f_{\text{U},2}\left(\varepsilon \cdot p_{\text{L},1}^{(i-1)} ,\varepsilon \cdot p_{\text{L},3}^{(i-1)},\varepsilon \cdot p_{\text{L},2}^{(i-1)}\right)\\
p_{\text{U},3}^{(i)}=&f_{\text{U},3}\left(\varepsilon \cdot p_{\text{L},1}^{(i-1)} ,\varepsilon \cdot p_{\text{L},3}^{(i-1)},\varepsilon \cdot p_{\text{L},2}^{(i-1)}\right) \ , 
\end{align}
where $f_{\text{U},k}$ denotes the transfer function of $\text{T}_{\text{U}}$ for the $k$th connected edge.
Likewise, the DE equations for $\text{T}_{\text{L}}$ can be written by swapping $p_{\text{U},k}^{(i)}$ and $p_{\text{L},k}^{(i)}$ for $k=1,2,3$.

Similarly to PCCs and SCCs, the DE equations of a BCC form a vector recursion. 
In order to modify this recursion to scalar form, in the first step consider identical upper and lower factor nodes. 
It follows $f_{\text{U},k}=f_{\text{L},k}\triangleq f_k$ and $p_{\text{U},k}^{(i)}=p_{\text{U},k}^{(i)}\triangleq x_k$ for $k=1,2,3$.
Then we can rewrite the DE equations of $\text{T}_{\text{U}}$ as
\begin{align}
\label{BCC1}
x_1^{(i+1)}&=f_1\Big(\varepsilon \cdot x_1^{(i)},\varepsilon \cdot x_3^{(i)},\varepsilon\cdot x_2^{(i)}\Big)\\
\label{BCC2}
x_2^{(i+1)}&=f_2\Big(\varepsilon \cdot x_1^{(i)},\varepsilon \cdot x_3^{(i)},\varepsilon \cdot x_2^{(i)}\Big)\\
\label{BCC3}
x_3^{(i+1)}&=f_3\Big(\varepsilon \cdot x_1^{(i)},\varepsilon \cdot x_3^{(i)},\varepsilon \cdot x_2^{(i)}\Big).
\end{align}
According to the above equations, the DE recursion is still in vector form. 
To rewrite it in scalar form, one alternative is to consider identical component encoders with a time-varying trellis, such that all three transfer functions are equal. 
For example, by periodically changing the order of symbols along trellis branches, this function becomes the average of the transfer functions $f_1,f_2,f_3$, $f_{\text{ave}}=\frac{f_1+f_2+f_3}{3}$.
By the above assumption, $x_1=x_2=x_3\triangleq x$.
Using this in \eqref{BCC1}-\eqref{BCC3} we can simplify the DE recursion as 
\begin{equation}
\label{eq:BCCScalar}
x^{(i+1)}=f_{\text{ave}}(\varepsilon \cdot x^{(i)},\varepsilon \cdot x^{(i)},\varepsilon \cdot x^{(i)}).
\end{equation}
Considering $f(x;\varepsilon)=f_{\text{ave}}(\varepsilon \cdot x,\varepsilon \cdot x,\varepsilon \cdot x)$, $g(x)=x$ and Lemma  \ref{remark1}, \eqref{eq:BCCScalar} is the recursion of a scalar admissible system.

\section{Single System Potential}\label{Sec2}
\begin{figure}[t]
  \centering
    \includegraphics[width=0.8\linewidth]{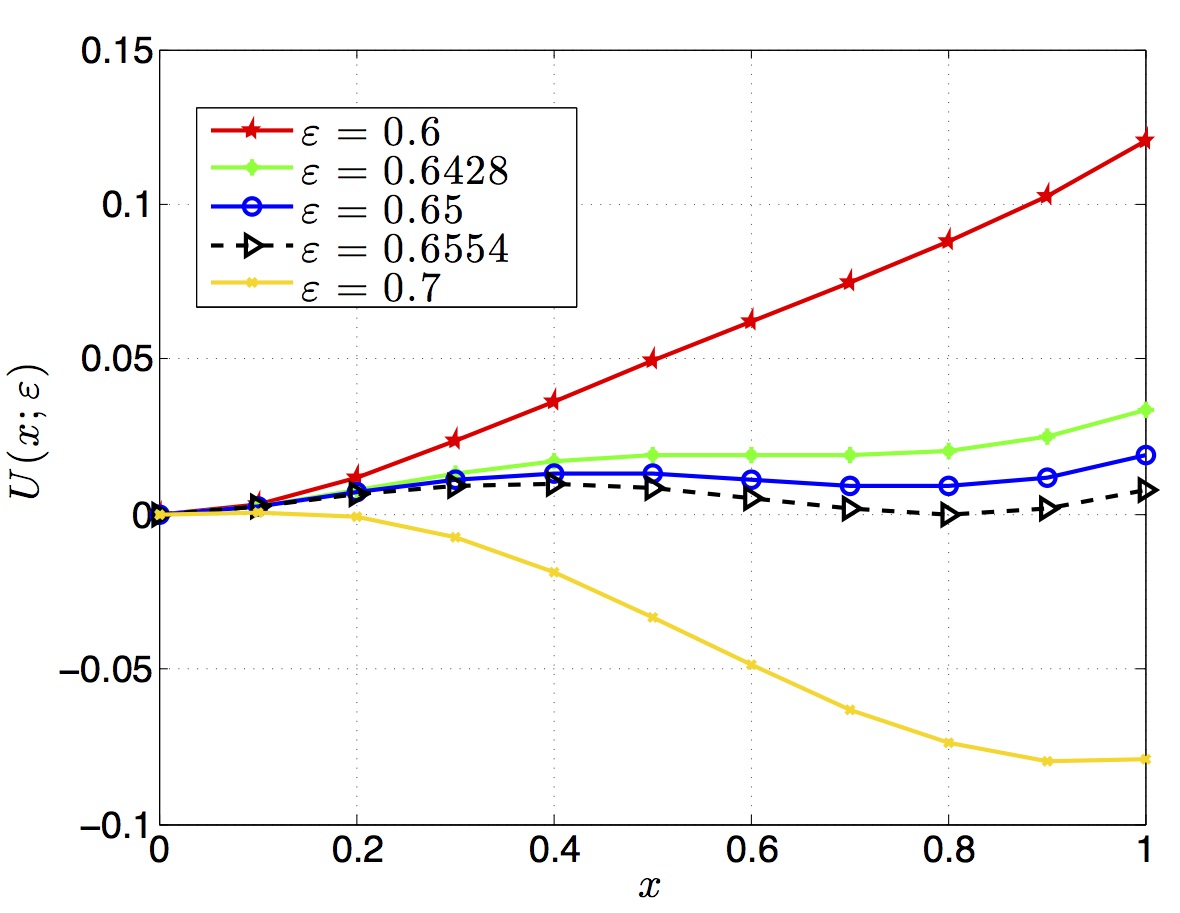}
\vspace{-2ex}
\caption{The potential function of PCC with generator matrix $\bs{G}=(1,5/7)$ in octal notation}
\label{PotPCC}
\vspace{-3ex}
\end{figure}

Since the DE recursion of TCs and BCCs can be written as the recursion of a scalar admissible system, we can derive the corresponding potential functions \cite{Yedla2012}.
\begin{definition} \label{def2}
For a scalar admissible system, the potential function $U(x;\varepsilon)$ is defined by
\begin{align}
\label{eq:Potential}
U(x;\varepsilon)&=\int_{0}^{x}\big{(}z-f(g(x);\varepsilon)\big{)}g'(z)dz \nonumber\\
&=xg(x)-G(x)-F(g(x);\varepsilon),
\end{align}
where $F(x;\varepsilon)=\int_{0}^{x}f(z;\varepsilon) dz$ and $G(x)=\int_{0}^{x}g(z) dz$.
\end{definition}
\begin{proposition}
The potential function has the following characteristics.
\begin{itemize}
\item $U(x;\varepsilon)$ is strictly decreasing in $\varepsilon \in (0,1]$.
\item An $x\in [0,1]$ is a fixed point of the recursion
(\ref{recursion}) iff it is a stationary point of the potential function.
\end{itemize}
\end{proposition}
\begin{definition} \label{defBP} If the DE recursion is the recursion of a BP decoder, the BP threshold is \cite{Yedla2012}
\[
\varepsilon^{\text{BP}}=\sup\Big\{\varepsilon
  \in[0,1]|U'(x;\varepsilon)>0,\; \forall x\in (0,1]\Big\}.
\] 
\end{definition}
 According to Definition \ref{defBP},  for $\varepsilon < \varepsilon^{\text{BP}}$, the potential function has no stationary point and its derivative is always larger than zero for $x\in (0,1]$. 
\begin{definition}
\label{defMAP}
For $\varepsilon >\varepsilon^{\text{BP}} $, the minimum unstable fixed point is $u(\varepsilon)=\sup\big\{\tilde{x} \in [0,1]| f(g(x);\varepsilon)<x, x\in (0,\tilde{x})\big\}$.  Then, the potential threshold is \cite{Yedla2012}
\begin{align*}
\varepsilon^*=\sup \Big\{\varepsilon \in [0,1]|  u(x)>0, \min_{x \in [u(x),1]} U(x;\varepsilon)> 0 \Big\}.
\end{align*}
\end{definition}
The potential threshold depends on functions $g(x)$ and $f(x;\varepsilon)$.
Since at least one of these functions depends on the component encoders, $\varepsilon^*$ also depends on the component encoders.
\begin{example}
Consider a rate-1/3 PCC in Fig.~\ref{Fig1}(a) with identical component encoders with generator matrix $\bs{G}=(1,5/7)$ in octal notation.
Its potential function is 
\[
U(x;\epsilon)=x^2-G(x)-F_{\text{s}}(x;\epsilon)=\frac{x^2}{2}-F_{\text{s}}(x;\epsilon),
\] 
where $F_{\text{s}}(x;\varepsilon)=\int_{0}^{x}f_{\text{s}}(\varepsilon\cdot z,\varepsilon) dz$ and $G(x)=\int_{0}^{x}g(z)dz=\frac{x^2}{2}$.

The potential function of this code is shown in Fig.~\ref{PotPCC}.
As it is illustrated in the figure, $\varepsilon=0.6428$ is the maximum channel erasure probability for which the derivative of the potential function is greater than zero and the potential function has no stationary point for $x\in (0,1]$. 
Thus,  $\varepsilon=0.6428$ is the BP threshold of this code (see Definition \ref{defBP}). 
The potential threshold is $\varepsilon^*=0.6554$ (see the black line in Fig.~ \ref{PotPCC}).
These results match with our numerical results in \cite{MoloudiISTC14}.
\end{example}

\section{Coupled system and Threshold Saturation} \label{Sec3}
\begin{theorem}\label{Thm:SC}
Consider a spatially coupled system defined by the following recursion at time $t$,
\begin{align}
\label{SC}
x_t^{(i+1)}=\frac{1}{1+m}\sum_{k=0}^{m}f\Big(\frac{1}{1+m}\sum_{j=0}^{m}g(x_{t+j-k}^{(i)});\varepsilon\Big).
\end{align}
For a large enough coupling memory and $\varepsilon < \varepsilon^*$, the only fixed point of the recursion is
$x=0$.
\end{theorem}

\begin{IEEEproof}
The proof follows from \cite{Yedla2012}.
\end{IEEEproof}

In the following we demonstrate that the recursion of  SC-PCCs, SC-SCCs and BCCs correspond to the recursion in (\ref{SC}).
\subsection{Spatially Coupled Parallel Concatenated Codes}
\begin{figure}[t]
  \centering
    \includegraphics[width=0.75\linewidth]{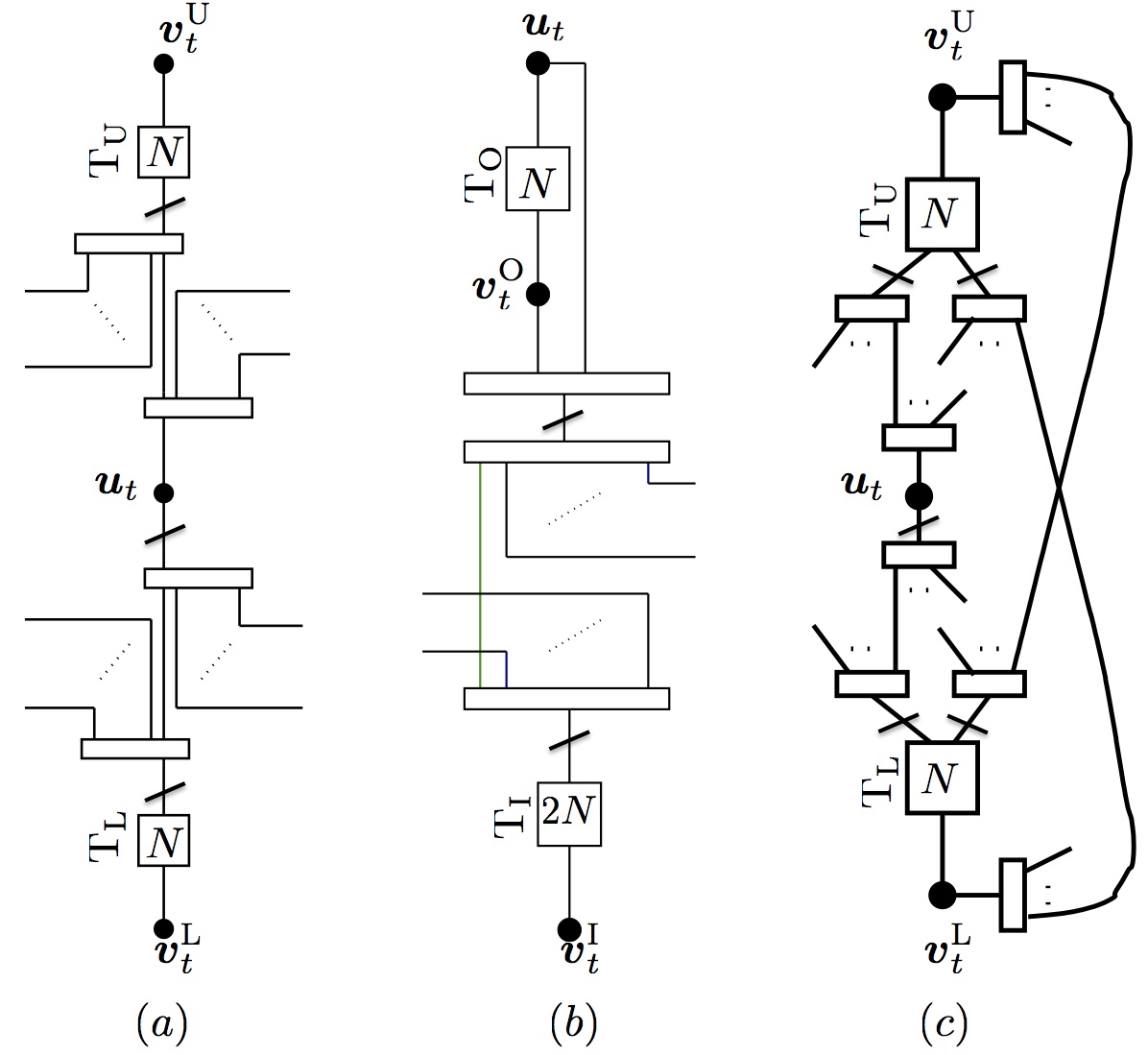}
\vspace{-2ex}
\caption{Compact graph of (a) SC-PCCs, (b) SC-SCCs, (c) BCCs}
\label{SCsystems}
\vspace{-4ex}
\end{figure}

Fig.~\ref{SCsystems}(a) shows the compact graph of SC-PCCs at time $t$ for coupling memory $m$. 
The sequence corresponding to the input variable node at this time, $\boldsymbol{u}_{t}$, is divided into $m+1$ sequences, $\boldsymbol{u}_{t,j}$, $j=0,\dots ,m$.
At time $t$, the sequences $\bs{u}_{t-j,j}$, $j=0,\dots ,m$ are multiplexed and reordered.
 The resulting sequence is used as the input to $\text{T}_{\text{U}}$ at time $t$. 
Likewise, a reordered copy of the sequence corresponding to the input variable node at the current time slot, $\boldsymbol{u}'_{t}$, is divided into $m+1$ sequences $\boldsymbol{u}'_{t,j'}$, $j'=0,\dots ,m$. 
At time $t$, the sequences $\boldsymbol{u}_{t-j',j'}$, $j'=0,\dots ,m$ are multiplexed and reordered. 
The resulting sequence is used as the input to $\text{T}_{\text{L}}$ at time $t$.
In other words, $\boldsymbol{u}_{t}$ is connected to the set of $\text{T}_{\text{U}}$s and the set of $\text{T}_{\text{L}}$s at time slots $t$ to $t+m$.
Consider identical $\text{T}_{\text{U}}$s and $\text{T}_{\text{L}}$s.
Due to the symmetric coupling structure, both erasure probabilities that come to $\boldsymbol{u}_{t}$ at time $t$ are equal and denoted by $x^{(i)}_{t}$. 
Following the compact graph of SC-PCCs, the erasure probability to $\text{T}_{\text{U}}$ through its first edge is the average of the erasure probabilities from $\boldsymbol{u}_{t'}$, $t'=t-m, \dots, t$. 
Therefore, the update of $\text{T}_{\text{U}}$ and $\text{T}_{\text{L}}$ at time $t$ is
\[
f_{\text{s},t}\Big(\frac{\varepsilon}{m+1}\cdot\sum_{j=0}^{m}x_{t-j}^{(i)},\varepsilon\Big),
\]
where $f_{\text{s},t}$ is the transfer function of $\text{T}_{\text{U}}$ and $\text{T}_{\text{L}}$ at time $t$ for the information bits. 
The erasure probability that comes to $\boldsymbol{u}_{t}$ at time slot $t$ through each of the incoming edges is the average of the erasure probabilities that come from the set of $\text{T}_{\text{U}}$s or $\text{T}_{\text{L}}$s at time slots $t$ to $t+m$. 
The recursion equation at time slot $t$ can then be written as
\begin{equation}
\label{eq:SCPCC}
x_t^{(i+1)}=\frac{1}{1+m}\sum_{k=0}^{m}f_{\text{s},t+k}\Big(\frac{\varepsilon}{m+1}\cdot\sum_{j=0}^{m}x_{t-j+k}^{(i)},\varepsilon\Big).
\end{equation}
The recursion (\ref{eq:SCPCC}) is identical to the recursion in (\ref{SC}). 
Thus, according to Theorem \ref{Thm:SC}, for channels with erasure probability $\varepsilon<\varepsilon^*$, the only fixed point of recursion \eqref{eq:SCPCC} is zero.

\subsection{Spatially Coupled Serially Concatenated Codes}
Fig.~\ref{SCsystems}(b) shows the compact graph of SC-SCCs at time $t$ for coupling memory $m$. 
Similarly to uncoupled SCCs, at time $t$, $\boldsymbol{u}_{t}$ and $\boldsymbol{v}_{t}^{\text{O}}$ are multiplexed and reordered to produce the sequence $\tilde{\boldsymbol{v}}_{t}^{\text{O}}$.
$\tilde{\boldsymbol{v}}_{t}^{\text{O}}$ is randomly divided into $m+1$ sequences $\tilde{\boldsymbol{v}}_{t,j}^{\text{O}}$, $j=0,\dots,m$.
$\text{T}_{\text{I}}$ at time $t$ receives a sequence which is built from sequences $\tilde{\boldsymbol{v}}_{t-j,j}^{\text{O}}$, $j=0,\dots,m$, and reordered.

Consider identical $\text{T}_{\text{O}}$ and $\text{T}_{\text{I}}$  and denote the erasure probability from $\text{T}_{\text{I}}$ to the set of $\boldsymbol{u}_{t}$ and $\boldsymbol{v}_{t}^{\text{O}}$ by $x^{i}_{t}$ at time $t$ and iteration $i$. 
Then, the message from this set to the set of $\text{T}_{\text{I}}$s is $g(x^{(i)}_{t})$, where $g(x)$ is obtained in (\ref{eq:gSCC}). 
Following the edges which are connected to $\text{T}_{\text{I}}$ at time $t$, the erasure probability that $\text{T}_{\text{I}}$ receives through its first edge is
\[
\frac{\varepsilon}{m+1}\cdot\sum_{j=0}^{m}g(x_{t-j}^{(i)}).
\]
The update of $\text{T}_{\text{I}}$ can be written as
\[
f_{\text{s},t}\Big(\frac{\varepsilon}{m+1}\cdot\sum_{j=0}^{m}g(x_{t-j}^{(i)}),\varepsilon\Big),
\]
where $f_{\text{s},t}$ is the transfer function of $\text{T}_{\text{I}}$ and $\text{T}_{\text{O}}$ at time $t$ for their input bits.

Both $\boldsymbol{u}_{t}$ and $\boldsymbol{v}_{t}^{\text{O}}$ receive equal erasure probabilities from the set of $\text{T}_{\text{I}}$s.
 This erasure probability is the average of the erasure probabilities from $\text{T}_{\text{I}}$ at time slots $t$ to $t+m$.
The DE recursion can then be written as
\begin{equation*}
\label{eq:SCSCC}
x_t^{(i+1)}=\frac{1}{1+m}\sum_{k=0}^{m}\varepsilon \cdot f_{\text{s},t+k}\Big(\frac{\varepsilon}{m+1}\cdot\sum_{j=0}^{m}g(x_{t-j+k}^{(i)}),\varepsilon\Big),
\end{equation*}
where  function $g$ is given in (\ref{eq:gSCC}).
This recursion is identical to the recursion in (\ref{SC}).
Therefore, we can conclude that for $\varepsilon<\varepsilon^*$ the only fixed point of the recursion of SC-SCCs is zero and threshold saturation occurs.
\subsection{Braided Convolutional Codes}
Fig.~\ref{SCsystems}(c) shows the compact graph of BCCs at time $t$ for coupling memory $m$.
This ensemble of BCCs is slightly different from the ensembles we introduced in \cite{Moloudi_SPCOM14}.
To couple the code with memory $m$, each of the sequences corresponding to $\bs{u}_t$, a reordered copy of the information sequence, $\tilde{\bs{u}}_t$, $\bs{v}_{t}^{\text{U}}$ and $\bs{v}_{t}^{\text{L}}$ is divided into $m+1$ sequences and denoted by $\bs{u}_{t,j}$, $\tilde{\bs{u}}_{t,j}$, $\bs{v}_{t,j}^{\text{U}}$ and $\bs{v}_{t,j}^{\text{L}}$ for $j=0, \dots, m$, respectively.
At time $t$, sequences $\bs{u}_{t-j,j}$ for, $j=0, \dots, m$, are multiplexed and reordered. 
The resulting sequence is used as the first input to $\text{T}_{\text{U}}$. 
Likewise, the sequences $\bs{v}_{t-j,j}^{\text{L}}$ for $j=0, \dots, m$, are multiplexed and reordered. The resulting sequence is used as the second input of $\text{T}_{\text{U}}$.
The sequences $\tilde{\bs{u}}_{t,j}$ for $j=0, \dots, m$, are multiplexed and reordered and used as the first input of $\text{T}_{\text{L}}$.
Likewise, the sequences $\bs{v}_{t-j,j}^{\text{U}}$ for $j=0, \dots, m$, are multiplexed and reordered and the resulting sequence is used as second input of $\text{T}_{\text{L}}$.

Consider identical component encoders at time $t$. The erasure probabilities to  $\bs{u}_t$, $\bs{v}_{t}^{\text{U}}$ and $\bs{v}_{t}^{\text{L}}$ are equal due to the symmetric coupling structure and denoted by $x^{(i)}_{t}$. 
Following the compact graph, the erasure probabilities to $\text{T}_{\text{U}}$ through all its incoming edges are equal and are given by the average of the erasure probabilities from $\bs{u}_{t'}$s, $t'= t-m,\dots,t$, $
q_t=\frac{\varepsilon}{1+m}\sum_{i=0}^{m}x_{t-j}^{(i)}$.
Thus, the erasure probabilities from each of the factor nodes to their outgoing edges are equal to $f_{\text{ave},t}(q_t,q_t,q_t),$
where $f_{\text{ave},t}$ is the transfer function of $\text{T}_{\text{U}}$ and $\text{T}_{\text{L}}$ at time $t$ for all edges. 
Finally the recursion at time slot $t$ is
\begin{equation}
\label{eq:SCBCC}
x_t^{(i+1)}=\frac{1}{1+m}\sum_{k=0}^{m}f_{\text{ave},t+k}(q_{t+k},q_{t+k},q_{t+k}).
\end{equation}

As (\ref{eq:SCBCC}) is identical to (\ref{SC}), according to Theorem \ref{Thm:SC}, for channels with erasure probability $\varepsilon<\varepsilon^*$, the only fixed point of (\ref{eq:SCBCC}) is equal to zero.

\section{Conclusions}
We considered three families of spatially-coupled turbo-like codes with identical component encoders whose density evolution recursions can be analyzed using the coupled scalar recursion framework of \cite{Yedla2012}. Then, based on this framework, we proved threshold saturation for these code ensembles over the BEC. For a more general case (different component encoders), the analysis is significantly more complicated and requires the coupled vector recursion framework of \cite{Yedla2012Vector}.



\begin{thebibliography}{10}

\bibitem{JimenezLDPCCC}
{A. {Jim{\'e}nez} Feltstr{\"o}m} and {K.Sh. Zigangirov},
\newblock ``Periodic time-varying convolutional codes with low-density
  parity-check matrices,''
\newblock {\em {IEEE} Trans. Inf. Theory}, vol. 45, no. 5, pp. 2181--2190,
  Sept. 1999.

\bibitem{Kudekar_ThresholdSaturation}
S.~Kudekar, T.J. Richardson, and R.L. Urbanke,
\newblock ``Threshold saturation via spatial coupling: {W}hy convolutional
  {LDPC} ensembles perform so well over the {BEC},''
\newblock {\em {IEEE} Trans. Inf. Theory}, vol. 57, no. 2, pp. 803 --834, Feb.
  2011.

\bibitem{ZhangBCC}
{W. Zhang}, {M. Lentmaier}, {K.Sh. Zigangirov}, and {D.J. Costello, Jr.},
\newblock ``Braided convolutional codes: a new class of turbo-like codes,''
\newblock {\em {IEEE} Trans. Inf. Theory}, vol. 56, no. 1, pp. 316--331, Jan.
  2010.

\bibitem{MoloudiISTC14}
S.~Moloudi, M.~Lentmaier, and A.~Graell~i Amat,
\newblock ``Spatially coupled turbo codes,''
\newblock in {\em Proc. Int. Symp. on Turbo Codes and Iterative Inform.}, Aug.
  2014.

\bibitem{BerrouTC}
C.~Berrou, A.~Glavieux, and P.~Thitimajshima,
\newblock ``Near {Shannon} limit error-correcting coding and decoding:
  turbo-codes,''
\newblock in {\em Proc.\ {IEEE} International Conference on Communications},
  Geneva, Switzerland, May 1993.

\bibitem{Benedetto98Serial}
S.~Benedetto, D.~Divsalar, G.~Montorsi, and F.~Pollara,
\newblock ``Serial concatenation of interleaved codes: performance analysis,
  design, and iterative decoding,''
\newblock {\em {IEEE} Trans. Inf. Theory}, vol. 44, no. 3, pp. 909--926, May
  1998.

\bibitem{MoloudiISIT14}
S.~Moloudi and M.~Lentmaier,
\newblock ``Density evolution analysis of braided convolutional codes on the
  erasure channel,''
\newblock in {\em Proc. IEEE Int. Symp. Inf. Theory}, Honolulu, HI, USA, July
  2014.

\bibitem{Moloudi_SPCOM14}
S.~Moloudi, M.~Lentmaier, and A.~Graell~i Amat,
\newblock ``Braided convolutional codes - a class of spatially coupled
  turbo-like codes,''
\newblock in {\em Proc.~Int. Conf. on Signal Processing and Communications},
  Bangalore, India, July 2014.

\bibitem{Moloudi_ISTW14}
A.~Graell~i Amat, S.~Moloudi, and M.~Lentmaier,
\newblock ``Spatially coupled turbo codes: Principles and finite length
  performance,''
\newblock in {\em Proc. Int. Symp. on Wireless Communications Systems}, Aug.
  2014.

\bibitem{Yedla2012}
A.~Yedla, Yung-Yih Jian, P.S. Nguyen, and H.D. Pfister,
\newblock ``A simple proof of threshold saturation for coupled scalar
  recursions,''
\newblock in {\em Proc. Int. Symp. on Turbo Codes and Iterative Inform.}, Aug.
  2012.

\bibitem{Yedla2012Vector}
A.~Yedla, Yung-Yih Jian, P.S. Nguyen, and H.D. Pfister,
\newblock ``A simple proof of threshold saturation for coupled vector
  recursions,''
\newblock in {\em Proc. IEEE Inform. Theory Workshop}, Sept. 2012.

\bibitem{Loeliger}
F.R. Kschischang, B.J. Frey, and H.-A. Loeliger,
\newblock ``Factor graphs and the sum-product algorithm,''
\newblock {\em IEEE Trans. Inf. Theory}, vol. 47, no. 2, pp. 498--519, Feb.
  2001.

\end{thebibliography}

\end{document}